\documentclass[preprint,aps,amsmath,nofootinbib]{revtex4-1}
\usepackage{graphicx,array,dcolumn}
\usepackage{calc,tabularx, epsfig,mathrsfs}
\usepackage{hyperref}

\usepackage{amsmath,verbatim,enumerate}
\usepackage{amssymb}
\usepackage{wasysym}
\usepackage{xcolor}
\usepackage{multirow}
\usepackage{xspace}
\usepackage{slashed}
\allowdisplaybreaks[1]
\newlength{\figurewidth}
\newcommand{\beq}{\begin{equation}}
\newcommand{\eeq}{\end{equation}}
\newcommand{\bea}{\begin{eqnarray}}
\newcommand{\eea}{\end{eqnarray}}
\newcommand{\ba}{\begin{array}}
\newcommand{\ea}{\end{array}}

\newcommand{\mn}{{\mu\nu}}

\newcommand{\pt}{\partial}

\newcommand{\al}{\alpha}
\newcommand{\bt}{\beta}
\newcommand{\g}{\gamma}
\newcommand{\ep}{\epsilon}
\newcommand{\ta}{\theta}
\newcommand{\lam}{\lambda}
\newcommand{\Lam}{\Lambda}
\newcommand{\G}{\Gamma}

\newcommand{\de}{\delta}
\newcommand{\D}{\Delta}
\newcommand{\OM}{\Omega}
\newcommand{\om}{\omega}
\newcommand{\sg}{\sigma}

\makeatother
\begin{document}
%
\title{
Surprises in Lorentzian path-integral of Gauss-Bonnet gravity
}
\setlength{\figurewidth}{\columnwidth}
%
\author{Gaurav Narain$\,{}^a$}
\email{gaunarain@gmail.com}
\affiliation{${}^a$ 
Institute of Theoretical Physics, 
Faculty of Science, Beijing University of Technology (BJUT), Beijing 100124, China.
}

%
\begin{abstract}
In this paper we study the Lorentzian path-integral of Gauss-Bonnet gravity 
in the mini-superspace approximation in four spacetime dimensions
and investigate the transition amplitude from one configuration to another.
Past studies motivate us on 
imposing Neumann boundary conditions on initial boundary as they lead to stable 
behaviour of fluctuations. The transition amplitude is computed exactly 
while incorporating the non-trivial contribution coming from the Gauss-Bonnet 
sector of gravity. A saddle-point analysis involving usage of 
Picard-Lefschetz methods allow us to gain further insight 
of the nature of transition amplitude. 
Small-size Universe is Euclidean in nature which is shown by the 
exponentially rising wave-function. It reaches a peak after which 
the wave-function becomes oscillatory indicating an emergence of 
time and a Lorentzian phase of the Universe. We also notice an
interesting hypothetical situation when the wave-function of Universe 
becomes independent of the initial conditions completely, 
which happens when cosmological constant 
and Gauss-Bonnet coupling have a particular relation. 
This however doesn't imply that the initial momentum is left
arbitrary as it needs to be fixed to a particular 
value which is chosen by demanding regularity of Universe at 
an initial time and the stability of fluctuations.

\end{abstract}

\maketitle

\tableofcontents

%

\section{Introduction}
\label{intro}

General relativity (GR) is highly successful in explaining a variety of physical phenomenon 
ranging from astrophysical to cosmological scales.
However, this model of gravitational theory lacks a high-energy 
completion and loses its reliability at short distances
\cite{tHooft:1974toh,Deser:1974nb,Deser:1974cz,Deser:1974cy,
Goroff:1985sz,Goroff:1985th,vandeVen:1991gw}.
Similarly, at ultra large scales its theoretical predictions 
don't agree with observational data, and one has to invoke 
dark-matter and/or dark energy to make an attempt at explaining them.
Following these various models have been proposed to amend GR 
at such extreme scales.  

Noticing a lack of renormalizability of GR 
(which has only two time derivatives of the metric field) one is motivated 
to modify GR at high-energies by incorporating 
higher-time derivatives of the metric field.
Such higher-derivative amendments although tackles 
issues of renormalizabilty but the theory develops 
another problem: lack of unitarity \cite{Stelle:1976gc,Salam:1978fd,Julve:1978xn}
(see also \cite{Fradkin:1981iu, Avramidi:1985ki,Buchbinder:1992rb} for some earlier works 
on higher-derivative gravity).
Some efforts have been made to deal with these issues 
in \cite{Narain:2011gs,Narain:2012nf,Narain:2017tvp,Narain:2016sgk}, 
in asymptotic safety approach 
\cite{Codello:2006in,Niedermaier:2009zz}
and `\textit{Agravity}' \cite{Salvio:2014soa}. 

The Gauss-Bonnet (GB) gravity in four spacetime dimensions is one such 
simple modification of the GR, where 
the highest order of time derivative of the metric-field remains two
and issues of \textit{ghosts} don't arise. 
Moreover, the additional term in the GB-gravity action 
is topological in four spacetime dimensions. Its addition doesn't 
change the dynamical evolution of spacetime metric.
However, it has a role to play in classifying topologies 
in path-integral quantization 
and hence has a non-trivial role to play at the boundaries
of manifolds. 
The Gauss-Bonnet gravity action is following
\bea
\label{eq:act}
S = \frac{1}{16\pi G} \int {\rm d}^Dx \sqrt{-g}
\biggl[
-2\Lam + R + \al
\biggl( R_{\mu\nu\rho\sg} R^{\mu\nu\rho\sg} - 4 R_\mn R^\mn + R^2 \biggr)
\biggr] \, , 
\eea
where $G$ is the Newton's gravitational constant, 
$\Lam$ is the cosmological constant term,  
$\al$ is the Gauss-Bonnet (GB) coupling and $D$ is spacetime dimensionality. 
The mass dimensions of various couplings are: 
$[G] = M^{2-D}$, $[\Lam] = M^2$ and $[\al] = M^{-2}$. 

This action falls in the class of lovelock gravity theories 
\cite{Lovelock:1971yv,Lovelock:1972vz,Lanczos:1938sf}, and
are a special class of higher-derivative gravity 
where equation of motion for the metric field remains second order in time.
Interestingly, GB term also arises in the
low-energy effective action of the heterotic string theory 
\cite{Zwiebach:1985uq,Gross:1986mw,Metsaev:1987zx},
and for the first time the coupling $\al$ has received observational constraints
\cite{Chakravarti:2022zeq}. These constraints come from the analysis 
of the gravitational wave (GW) data of the event 
GW150914 which also offered the first observational 
confirmation of the area theorem \cite{Isi:2020tac}.

My interest in this paper is to investigate the path-integral of the 
gravity where the gravitational theory is given by the 
action in eq. (\ref{eq:act}), and study the effect of 
boundary conditions \cite{York:1986lje,Brown:1992bq,Krishnan:2016mcj,
Witten:2018lgb,Krishnan:2017bte}
on the wave-function of Universe. 
We start by considering a generic metric which is 
spatial homogenous and isotropic in $D$ spacetime 
dimensions. It is the FLRW metric in arbitrary spacetime 
dimension with dimensionality $D$. In polar
co-ordinates $\{t_p, r, \ta, \cdots \}$ it is given by
\beq
\label{eq:frwmet}
{\rm d}s^2 = - N_p^2(t_p) {\rm d} t_p^2 
+ a^2(t_p) \left[
\frac{{\rm d}r^2}{1-kr^2} + r^2 {\rm d} \OM_{D-2}^2
\right] \, .
\eeq
It has two unknown time-dependent functions: lapse $N_p(t_p)$
and scale-factor $a(t_p)$, $k=(0, \pm 1)$ is the curvature, 
and ${\rm d}\OM_{D-2}$ is the 
metric for the unit sphere in $D-2$ spatial dimensions. 
This is the \textit{mini-superspace} approximation of the metric.
This is a huge simplification of the original gravitational theory 
in a sense as we do not have have gravitational waves
anymore in this reduced framework.
However, we still do retain diffeomorphism invariance of the time 
co-ordinate $t_p$ and the dynamical scale-factor $a(t_p)$. 
This simple setting is enough for exploring 
issues of gravitational path-integral involving boundary 
conditions where GB-modifications can/may play a non-trivial role.

The Feynman path-integral for the theory in reduced space can be written as
\beq
\label{eq:Gform_sch_fpt}
G[{\rm bd}_0, {\rm bd}_1]
= \int_{{\rm bd}_0}^{{\rm bd}_1} {\cal D} N_p {\cal D} \pi {\cal D} 
a(t_p) {\cal D} p {\cal D} {\cal C} {\cal D} \bar{P}
\exp \biggl[
\frac{i}{\hbar} \int_0^1 {\rm d} t_p \left(
N^\prime_p \pi + a^\prime p +{\cal C}^\prime \bar{P} - N_p H \right)
\biggr] \, ,
\eeq
where beside the scale-factor $a(t_p)$, lapse $N_p$ and fermionic ghost
${\cal C}$, we also have their corresponding conjugate momenta given by
$p$, $\pi$ and $\bar{P}$ respectively. And once again the 
$({}^\prime)$ here denotes derivative 
with respect to $t_p$. The original path-integral measure then changes to a 
measure over all these variables. Without loss of generality one choose the 
time $t_p$ co-ordinate to range from $0\leq t_p \leq 1$. 
Here ${\rm bd}_0$ and ${\rm bd}_1$ are the field configuration at 
initial ($t_p=0$) and final ($t_p=1$) boundaries respectively. 
The Hamiltonian constraint $H$ consists of two parts 
\beq
\label{eq:Htwo}
H = H_{GB}[a, p] + H_{\rm gh} [N, \pi, {\cal C}, \bar{P}] \, ,
\eeq
where $H_{GB}$ refers to the Hamiltonian corresponding to Gauss-Bonnet gravity action 
and the Batalin-Fradkin-Vilkovisky (BFV) \cite{Batalin:1977pb}
ghost Hamiltonian is denoted by $H_{\rm gh}$
\footnote{ 
The BFV ghost is a generlization of the usual Fadeev-Popov ghost which is 
based on BRST symmetry. In standard gauge theories the 
constraint algebra forms a Lie algebra. However, the constraint algebra 
doesn't closes in case of gravitational theories which respect diffeomorphism invariace. 
For this reason one needs BFV quantization process. In 
mini-superspace approximation there is only one constraint, which is the 
Hamiltonian $H$. Here the algebra therefore trivially closes leaving the 
distinction between two quantization process irrelevant. 
Nevertheless BFV quantization is still preferable.}.
In the mini-superspace approximation we still have some diffeomorphism invariance 
left which shows up as a time reparametrization symmetry. 
In order to break this invariance we do gauge-fixing by 
choosing $N^\prime_p=0$ (proper-time gauge).
For more elaborate discussion on BFV quantization process and ghost
see \cite{Teitelboim:1981ua,Teitelboim:1983fk,Halliwell:1988wc}. 

In the mini-superspace approximation most of the path-integral in 
eq. (\ref{eq:Gform_sch_fpt}) can be performed analytically leaving behind 
the following path-integral 
\beq
\label{eq:Gform_sch}
G[{\rm bd}_0, {\rm bd}_1]
= \int_{0^+}^{\infty} {\rm d} N_p
\int_{{\rm bd}_0}^{{\rm bd}_1} {\cal D} a(t_p) \,\,
e^{i S[a, N_p]/\hbar} \, .
\eeq
This is easy to interpret as the path-integral 
$\int {\cal D} a(t_p) \,\, e^{i S[a, N_p]/\hbar}$ represent the quantum-mechanical 
transitional amplitude for the Universe to evolve from one 
field configuration to another in the proper time $N_p$.
The lapse-integration $N_p$ indicates that one need to consider paths 
of every proper duration $0<N_p<\infty$.
This choice implies causal evolution from 
one field configuration ${\rm bd}_0$ to another ${\rm bd}_1$ 
as shown in \cite{Teitelboim:1983fh}, where $a_0<a_1$ will refer to expanding 
Universe while $a_0>a_1$ will imply contracting Universe. 

In this paper we are interested in investigating this 
path-integral more carefully for the case of Gauss-Bonnet gravity
where we study the effects of boundary conditions 
and the non-trivial manner it affects the path-integral 
when GB modifications of gravity are taken into account.
In principle the boundary configurations
are chosen in such a way so that the variational problem leading to 
equation of motion (and its solution) are consistent, but it is 
important (and actually better) to choses the ones which lead to 
stable perturbations around the saddle points. In this case the 
path-integral becomes a summation over all the \textit{stable} 
geometries, where boundary conditions leading to 
unstable saddle points are not incorporated. It is a 
\textit{stability} condition.

Generically, such path-integrals require to be analysed 
systematically in a framework of complex 
analysis. Recent work \cite{Kontsevich:2021dmb} on the `allowability' criterion 
provides a simple diagnostic-tool for identifying the 
allowable complex metrics 
on which quantum field theories can be consistently defined. 
It is still to be seen whether this \textit{criterion} is 
necessary or sufficient (see also recent works 
\cite{Witten:2021nzp} and \cite{Lehners:2021mah}).
Our approach is to avoid performing Wick-rotation to 
Euclidean signature at all and aim to tackle the gravitational 
path-integral directly in Lorentzian signature itself. 

Picard-Lefschetz theory offers a process to carefully handle 
such oscillatory path-integrals in a systematic manner. 
It provides a framework where Lorentzian, complex and Euclidean saddle points 
can be treated democratically. It is an extension of the 
standard Wick-rotation prescription
to define convergent 
contour integral on a generic curved spacetime
\footnote{
Some attempts to do Wick-rotation sensibly in curved spacetime have been
made in \cite{Candelas:1977tt,Visser:2017atf,Baldazzi:2019kim,Baldazzi:2018mtl}.
However, more work needs to be done in this.
}. 
This framework allows one to uniquely determine 
contours of integrations along which the integrands like the 
ones appearing in eq. (\ref{eq:Gform_sch}) are well-behaved.
By definition then the original oscillatory integrals become 
convergent along these contours which are 
termed \textit{Lefschetz thimbles}. This framework has been recently 
used in the last few years to probe issues in Lorentzian
quantum cosmology \cite{Feldbrugge:2017kzv,Feldbrugge:2017fcc,Feldbrugge:2017mbc,
Vilenkin:2018dch, Vilenkin:2018oja, Rajeev:2021xit}
and study effects of various boundary conditions \cite{DiTucci:2019dji,DiTucci:2019bui,
Narain:2021bff,Lehners:2021jmv}
\footnote
{
Earlier attempts using complex analysis were made in
studying Euclidean gravitational path-integrals 
which are known to suffer 
from conformal factor problem \cite{Hawking:1981gb,Hartle:1983ai}.
In the context of Euclidean quantum cosmology the usage of complex analysis 
was made to explore issues regarding initial conditions: 
\textit{tunnelling} proposal \cite{Vilenkin:1982de,Vilenkin:1983xq,Vilenkin:1984wp} 
and 
\textit{no-boundary} proposal \cite{Hawking:1981gb,Hartle:1983ai,Hawking:1983hj}.
Beside the initial conditions one also need a sensible choice 
of integration contour to have path-integral 
well-defined \cite{Halliwell:1988ik,Halliwell:1989dy,Halliwell:1990qr}.
}.

Once we have a way to define the oscillatory path-integral 
in a systematic fashion, we are then in a position to explore the 
consequences of the various boundary conditions and 
determine the favourable ones by analysing the behaviours 
of perturbations. Past studies aimed at investigating the 
no-boundary proposal of Universe in the context of Lorentzian 
quantum gravity have investigated Dirichlet 
boundary conditions(DBC) \cite{Feldbrugge:2017kzv,Feldbrugge:2017fcc,Feldbrugge:2017mbc},
Neumann boundary conditions (NBC) \cite{DiTucci:2019bui,Narain:2021bff,
Lehners:2021jmv, DiTucci:2020weq}, robin boundary conditions (RBC) 
\cite{DiTucci:2019dji,DiTucci:2019bui}. It is seen that in 
DBC the perturbations around the \textit{relevant} complex-saddle 
point are not suppressed resulting them being disfavoured, 
while this don't happen in case of NBC and RBC
where the fluctuations around the \textit{relevant} saddles are suppressed
(see \cite{Feldbrugge:2017kzv, Narain:2021bff} for a concise 
review on Picard-Lefschetz theory and process of determining 
\textit{relevance/irrelevance} of saddle points). These studies 
show that in order to have a well-defined no-boundary proposal 
of Universe one should make use of Neumann (or Robin) BC 
either at initial boundary or final boundary or at both boundaries. These 
studies further support the simple situation where Neumann BC 
is imposed at initial boundary while a Dirichlet BC is imposed at 
final boundary \cite{DiTucci:2019bui,Narain:2021bff,DiTucci:2020weq, Lehners:2021jmv}, 
as the perturbations are well-behaved.
These results motivates us to investigate this particular situation 
of NBC more carefully and apply it to the Lorentzian 
path-integral of gravity where the gravitational action is given
by the action in eq. (\ref{eq:act}).

The outline of paper is as follows: after an introduction in 
section \ref{intro}, we talk about the mini-superspace approximation 
and apply it to the gravitational theory 
in section \ref{minisup}. We then study the variational problem 
in section \ref{bound_act} and compute the boundary action needed 
to have a consistent variational problem. In section \ref{neumann}
we compute the boundary actions for the Neumann boundary condition 
at the initial boundary which allows to determine the total action 
of theory involving boundary terms. In section \ref{TranProb} we 
compute the expression for the transition amplitude and notice that 
it factors in two parts: one entirely dependent on initial boundary
and one entirely dependent on final boundary. In section 
\ref{airy} we compute the exact expression for the transition 
amplitude by making use of Airy-functions. Section \ref{sad_pot_approx}
is devoted to saddle-point analysis of the transition amplitude. 
In section \ref{inc_ind} we talk about a special scenario of 
initial condition independence. This is followed by 
conclusions in section \ref{conc}.

\section{Mini-superspace action}
\label{minisup}

The FLRW metric given in eq. (\ref{eq:frwmet}) is conformally-flat and hence
its Weyl-tensor $C_{\mu\nu\rho\sg} =0$. The non-zero 
entries of the Riemann tensor are 
\cite{Deruelle:1989fj,Tangherlini:1963bw,Tangherlini:1986bw} 
\bea
\label{eq:riemann}
R_{0i0j} &=& - \left(\frac{a^{\prime\prime}}{a} - \frac{a^\prime N_p^\prime}{a N_p} \right) g_{ij} \, , 
\notag \\
R_{ijkl} &=& \left(\frac{k}{a^2} + \frac{a^{\prime2}}{N_p^2 a^2} \right)
\left(g_{ik} g_{jl} - g_{il} g_{jk} \right) \, ,
\eea
where $g_{ij}$ is the spatial part of the FLRW metric
and $({}^\prime)$ denotes derivative with respect to $t_p$.
For the Ricci-tensor the non-zero components are 
\bea
\label{eq:Ricci-ten}
R_{00} &=& - (D-1) \left(\frac{a^{\prime\prime}}{a} - \frac{a^\prime N_p^\prime}{a N_p} \right)
\, , 
\notag \\
R_{ij} &=& \left[
\frac{(D-2) (k N_p^2 + a^{\prime2})}{N_p^2 a^2}
+ \frac{a^{\prime\prime} N_p - a^\prime N_p^\prime}{a N_p^3} 
\right] g_{ij} \, ,
\eea
while the Ricci-scalar for FLRW is given by
\beq
\label{eq:Ricci0}
R = 2(D-1) \left[\frac{a^{\prime\prime} N_p - a^\prime N_p^\prime}{a N_p^3} 
+ \frac{(D-2)(k N_p^2 + a^{\prime2})}{2N_p^2 a^2} \right]
\, .
\eeq
Moreover, for Weyl-flat metrics one can express 
Riemann tensor in terms of Ricci-tensor and Ricci scalar as follows
\bea
\label{eq:Riem_exp}
R_{\mu\nu\rho\sg} = \frac{R_{\mu\rho} g_{\nu\sg} - R_{\mu\sg}g_{\nu\rho}
+ R_{\nu\sg} g_{\mu\rho} - R_{\nu\rho} g_{\mu\sg}}{D-2}
- \frac{R (g_{\mu\rho} g_{\nu\sg} - g_{\mu\sg} g_{\nu\rho})}{(D-1)(D-2)} \,.
\eea
Usage of this identity implies that $R_{\mu\nu\rho\sg} R^{\mu\nu\rho\sg}$
can be written as follows
\beq
\label{eq:Reim2_exp}
R_{\mu\nu\rho\sg} R^{\mu\nu\rho\sg}
= \frac{4}{D-2} R_\mn R^\mn - \frac{2 R^2}{(D-1)(D-2)} \, .
\eeq
This when plugged in the $R_{\mu\nu\rho\sg} R^{\mu\nu\rho\sg}$  
in the GB-gravity action then we get the following for 
case of Weyl-flat metrics
\bea
\label{eq:actGB}
\int {\rm d}^Dx \sqrt{-g} && \left(
R_{\mu\nu\rho\sg} R^{\mu\nu\rho\sg}  - 4 R_\mn R^\mn + R^2
\right)
\notag \\
&&
= \frac{D-3}{D-2} \int {\rm d}^Dx \sqrt{-g} \left(
- 4 R_\mn R^\mn + \frac{D R^2}{D-1}
\right) \, .
\eea
On plugging the FLRW metric of eq. (\ref{eq:frwmet}) in the gravitational action 
stated in eq. (\ref{eq:act}), we get an action for scale-factor $a(t_p)$ and lapse
$N_p(t_p)$ in $D$-dimensions
\bea
\label{eq:midSact}
&&
S = \frac{V_{D-1}}{16 \pi G} \int {\rm d}t_p
\biggl[
\frac{a^{D-3}}{N_p^2} \biggl\{
(D-1)(D-2) k N_p^3 - 2 \Lam a^2 N_p^3 - 2 (D-1) a a^\prime N_p^\prime
\notag \\
&&
+ (D-1)(D-2) a^{\prime2} N_p + 2 (D-1) N_p a a^{\prime\prime}
\biggr\}
+ (D-1)(D-2)(D-3) \al\biggl\{
\frac{a^{D-5}(D-4)}{N_p^3} 
\notag\\
&&
\times (kN_p^2 + a^{\prime2})^2 
+ 4 a^{D-4}\frac{{\rm d}}{{\rm d}t_p} 
\left(
\frac{k a^\prime}{N_p} + \frac{a^{\prime 3}}{3N_p^3}
\right)
\biggr\}
\biggr] \, ,
\eea
where $V_{D-1}$ is the volume of $D-1$ dimensional sphere and is given by,
\beq
\label{eq:volDm1}
V_{D-1} = \frac{\G(1/2)}{\G(D/2)} \left(\frac{\pi}{k}\right)^{(D-1)/2} \, .
\eeq
An interesting thing happens in $D=4$ when 
the GB-sector terms proportional $\al$ 
becomes a total time-derivative. The mini-superspace 
gravitational action then becomes the following in $D=4$
\beq
\label{eq:mini_sup_d4}
S = \frac{V_{3}}{16 \pi G} \int {\rm d}t_p
\biggl[
6k a N_p - 2 \Lam a^3 N_p - \frac{6 a^2 a^\prime N_p^\prime}{N_p}
+ \frac{6 a a^{\prime 2}}{N_p} + \frac{6 a^{\prime\prime} a^2}{N_p}
+ 24 \al \frac{{\rm d}}{{\rm d}t_p} 
\left(
\frac{k a^\prime}{N_p} + \frac{a^{\prime 3}}{3N_p^3}
\right)
\biggr] \, .
\eeq
By a rescaling of lapse and scale-factor the above action can be 
recast into a more appealing form. If we do the following transformation 
\beq
\label{eq:rescale}
N_p(t_p) {\rm d} t_p = \frac{N(t)}{a(t)} {\rm d} t \, ,
\hspace{5mm}
q(t) = a^2(t) \, ,
\eeq
then our original FLRW metric in eq. (\ref{eq:frwmet})
changes into following
\beq
\label{eq:frwmet_changed}
{\rm d}s^2 = - \frac{N^2}{q(t)} {\rm d} t^2 
+ q(t) \left[
\frac{{\rm d}r^2}{1-kr^2} + r^2 {\rm d} \OM_{D-2}^2
\right] \, ,
\eeq
and our gravitational action in $D=4$ given in eq. (\ref{eq:mini_sup_d4}) 
acquires a following simple form
\bea
\label{eq:Sact_frw_simp}
S = \frac{V_3}{16 \pi G} \int_0^1 {\rm d}t \biggl[
(6 k - 2\Lam q) N + \frac{3 \dot{q}^2}{2N}
+ 3q \frac{{\rm d}}{{\rm d} t} \left(\frac{\dot{q}}{N} \right)
+ 24 \al \frac{{\rm d}}{{\rm d} t} \left(
\frac{k\dot{q}}{2N} + \frac{\dot{q}^3}{24 N^3} 
\right)
\biggr] \, .
\eea
Here $(\dot{})$ represent time $t$ derivative. 
With an integration by parts this action can be written in the following manner 
\bea
\label{eq:Sact_frw_simp_inp}
S &&
=\frac{V_3}{16 \pi G} \int_0^1 {\rm d}t \biggl[
(6 k - 2\Lam q) N - \frac{3 \dot{q}^2}{2N} \biggr] 
+ \frac{V_3}{16 \pi G} \biggl[
\frac{3q_1 \dot{q}_1}{N} - \frac{3q_0 \dot{q}_0}{N}
\notag \\
&&
+ 24 \al \left(
\frac{k\dot{q_1}}{2N} + \frac{\dot{q_1}^3}{24 N^3} 
- 
\frac{k\dot{q_0}}{2N} - \frac{\dot{q_0}^3}{24 N^3} 
\right)
\biggr]
\, ,
\eea
where we notice that there are two surface terms: one coming 
from EH-part of gravitational action while the other is GB term.
From now onwards we will work with the convention 
that $V_3 = 8\pi G$.

\section{Action variation and boundary terms}
\label{bound_act}

To find the equation of motion and construct a consistent variational problem
we start by considering the variation of the action in eq. (\ref{eq:Sact_frw_simp}) 
with respect to $q(t)$. From now onwards we will 
work in the ADM gauge $\dot{N}=0$, which implies 
setting $N(t) = N_c$ (constant). We write 
\beq
\label{eq:qfluc}
q(t) = \bar{q}(t) + \ep \de q(t)
\eeq
where $\bar{q}(t)$ satisfies the equation of motion, $\de q(t)$ is the fluctuation 
around this. The parameter $\ep$ is used to keep a track of the order of fluctuation terms. 
On plugging this in action in eq. (\ref{eq:Sact_frw_simp}) and on 
expanding it to first order in $\ep$ we have
\beq
\label{eq:Sexp_qvar}
\de S = \frac{\ep}{2} \int_{0}^{1} {\rm d}t \biggl[
\left(-2 \Lam N_c + \frac{3 \ddot{q}}{N_c} \right) \de q
+ \frac{3}{N_c} \frac{{\rm d}}{{\rm d} t} \left(q \de \dot{q} \right)
+ 24 \al \frac{{\rm d}}{{\rm d} t} \left\{ 
\left(\frac{k}{2N_c} + \frac{\dot{q}^2}{8N_c^3} \right) \de \dot{q} \right\}
\biggr] \, .
\eeq
We notice in the above that there are two total time-derivative pieces 
which becomes relevant at the boundaries and for consistent 
boundary value problem they need to be canceled appropriately by addition 
of suitable boundary actions. 
The term proportional to $\de q$ on the other hand 
gives the equation of motion for $q$
\beq
\label{eq:dyn_q_eq}
\ddot{q} = \frac{2}{3} \Lam N_c^2 \, .
\eeq
This linear second-order differential equation is easy to solve 
and its general solution is given by
\beq
\label{eq:qsol_gen}
q(t) = \frac{\Lam N_c^2}{3} t^2 + c_1 t + c_2 \, .
\eeq
Here $c_{1,2}$ are constants which gets determined based on the boundary conditions. 
The total-derivative terms in the above results in a collection of 
boundary terms
\beq
\label{eq:Sbd}
S_{\rm bdy} = \frac{\ep}{2} \biggl[
\frac{3}{N_c} \left(q_1 \de \dot{q}_1 - q_0 \de \dot{q}_0 \right)
+ 24 \al \left\{ 
\left(\frac{k\de \dot{q}_1}{2N_c} + \frac{\dot{q}_1^2\de \dot{q}_1}{8N_c^3} \right) 
-  \left(\frac{k\de \dot{q}_0}{2N_c} + \frac{\dot{q}_0^2\de \dot{q}_0}{8N_c^3} \right)\right\}
\biggr] \, ,
\eeq
where 
\beq
\label{eq:BC_name}
q_0 = q(t=0) \, , 
\hspace{5mm}
q_1 = q(t=1) \, ,
\hspace{5mm}
\dot{q}_0 = \dot{q}(t=0) \, ,
\hspace{5mm}
\dot{q}_1 = \dot{q}(t=1) \, .
\eeq
The constants $c_{1,2}$ will be fixed later depending on the 
choice of boundary conditions. The action given in eq. (\ref{eq:Sact_frw_simp_inp})
can be used to determine the conjugate momentum to the field $q$
\beq
\label{eq:mom_conju_qq}
\pi = \frac{\pt {\cal L}}{\pt \dot{q}} = - \frac{3\dot{q}}{2N_c} \, ,
\eeq
where we have used the ADM gauge. The above boundary terms can be 
written in terms of conjugate momentum as follows
\beq
\label{eq:Sbd_mom}
S_{\rm bdy} = - \ep\biggl[
\left(q_1 \de \pi_1 - q_0 \de \pi_0 \right)
+4 \al \left\{ 
\left(k\de \pi_1 + \frac{\pi_1^2\de \pi_1}{27} \right) 
-  \left(k\de \pi_0 + \frac{\pi_0^2\de\pi_0}{27} \right)\right\}
\biggr] \, .
\eeq
To cancel the boundary terms that arise 
during variation of action, one has to add surface terms in order to have a 
well-defined variational problem. In the present case this will mean that 
we supplement our original action given in eq. (\ref{eq:Sact_frw_simp})
with the following terms
\bea
\label{eq:Sact_surf_full}
&&
S_{\rm surface} 
= \frac{1}{2} \biggl[
-\left. \frac{3q \dot{q}}{N_c} \right|_0^1
- 24 \al \left. \left(
\frac{k\dot{q}}{2N_c} + \frac{\dot{q}^3}{24 N_c^3} 
\right) \right|_0^1
\biggr] 
\notag \\
&&
= 
(q_1 \pi_1 - q_0 \pi_0)
+ 4 \al \left(k\pi_1 +\frac{\pi_1^3}{27} 
- k\pi_0 - \frac{\pi_0^3}{27}\right)  \, .
\eea
Here the first term is Gibbon-Hawking-York (GHY) term 
\cite{York:1986lje,Gibbons:1978ac,Brown:1992bq}
imposed at the two boundaries, while the second term is a 
surface term needed to cancel the 
effects of GB at the two boundaries.

\section{Neumann Boundary condition (NBC) at $t=0$}
\label{neumann}

We notice that the variational problem can be made consistent if 
we impose Neumann boundary condition \cite{Krishnan:2016mcj,DiTucci:2019dji} 
at $t=0$ and a 
Dirichlet boundary condition at $t=1$. This will imply that
we have $\de q_1=0$ and $\de \dot{q}_0=0$ (or $\de \pi_0=0$). 
Imposing Neumann boundary conditions at $t=0$ is also favourable 
as it has been seen in past studies that the path-integral 
is well-behaved and the perturbations are suppressed 
\cite{DiTucci:2019bui,Narain:2021bff,DiTucci:2020weq, Lehners:2021jmv}.
This will mean
\beq
\label{eq:neuMa_cond}
\pi_0 \,\, \&  \,\,
q_1 = {\rm fixed} 
\hspace{5mm} 
\Rightarrow 
\hspace{5mm}
\de \pi_0 = 0  \hspace{3mm} \&  \hspace{3mm}
\de q_1 = 0 \, .
\eeq
This will imply that the boundary terms given in eq. (\ref{eq:Sbd_mom}) 
arising during the variation of action will reduce to the following 
\beq
\label{eq:Sbd_mom_NBC}
\left. S_{\rm bdy} \right|_{\rm NBC} = - \ep\biggl[
q_1 \de \pi_1
+4 \al 
\left(k\de \pi_1 + \frac{\pi_1^2\de \pi_1}{27} \right) 
\biggr] \, .
\eeq
In order to cancel these boundary terms and have a consistent variational 
problem one has to add the following surface term to our original action 
\beq
\label{eq:Sact_surf_NBC}
\left. S_{\rm surface} \right|_{\rm NBC}
= \frac{1}{2} \biggl[
-\frac{3q_1 \dot{q}_1}{N_c} 
- 24 \al \left(
\frac{k\dot{q}_1}{2N_c} + \frac{\dot{q}_1^3}{24 N_c^3} 
\right) 
\biggr] 
= q_1 \pi_1 
+ 4 \al \left(k\pi_1 +\frac{\pi_1^3}{27} \right)  \, .
\eeq
This means we introduce GHY-term and a Chern-Simon like term 
at the final boundary to have a consistent variational problem.
For these set of boundary conditions one can now determine 
$c_{1,2}$ in the solution to equation of motion for $q(t)$ 
given in eq. (\ref{eq:qsol_gen}). 
This will imply
\beq
\label{eq:qsol_nbc}
\bar{q}(t) = \frac{\Lam N_c^2}{3} (t^2-1) - \frac{2 N_c \pi_0}{3} (t-1) + q_1 \, ,
\eeq
where `bar' over $q$ is added as it is solution to equation of motion.
In this setting $t=0$ will give
\beq
\label{eq:q0_nbc}
q_0 = q_1 + \frac{2 N_c \pi_0}{3} - \frac{\Lam N_c^2}{3} \, .
\eeq
The surface terms can then be added to the action in eq. (\ref{eq:Sact_frw_simp_inp})
to obtain full action of the system. This is given by
\beq
\label{eq:Sact_frw_simp_nbc}
S_{\rm tot}
=\frac{1}{2} \int_0^1 {\rm d}t \biggl[
(6 k - 2\Lam q) N_c - \frac{3 \dot{q}^2}{2N_c} \biggr] 
+ \left(q_1 + \frac{2 N_c \pi_0}{3} - \frac{\Lam N_c^2}{3} \right) \pi_0
+ 4 \al \left(k\pi_0 +\frac{\pi_0^3}{27} \right)
\, ,
\eeq
where we have substituted the expression for $q_0$ using the equation 
(\ref{eq:q0_nbc}). Furthermore, if we substitute the solution to equation 
of motion eq. (\ref{eq:qsol_nbc}) in the above then we will obtain 
an on-shell action which is given by
\beq
\label{eq:stot_onsh_nbc}
S_{\rm tot}[\bar{q}] = \frac{\Lam^2}{9}  N_c^3 
- \frac{\Lam \pi_0}{3}  N_c^2 
+ \left(3k - \Lam q_1 + \frac{\pi_0^2}{3} \right) N_c
+ q_1 \pi_0 + 4\al \left(k\pi_0 + \frac{\pi_0^3}{27} \right) \, .
\eeq
This is also the action for the lapse $N_c$. 
Note that the action obtained when NBC is used is not singular at $N_c=0$,
which is not the case when DBC are used 
\cite{Feldbrugge:2017kzv,Feldbrugge:2017fcc,Feldbrugge:2017mbc,
DiTucci:2019bui,Narain:2021bff,
Lehners:2021jmv, DiTucci:2020weq}.
The point about the lack of $N_c=0$ singularity can be understood 
by realising that as we are fixing the initial momentum (and not the initial size of geometry),
we are therefore summing over all possible initial $3$-geometry size 
and their transition to $3$-geometry of size $q_1$. This summation 
will also include contribution of a transition from $q_1 \to q_1$. 
These transition can occur instantaneously {\it i.e.} with $N_c=0$,
thereby implying that there is nothing singular happening at $N_c=0$.

\section{Transition Amplitude}
\label{TranProb}

We are now in a position to ask about the transition amplitude
from one $3$-geometry to another. 
The relevant quantity that we wish to know can be expressed in mini-superspace approximation
as follows (see \cite{Halliwell:1988ik,Feldbrugge:2017kzv} for the Euclidean 
gravitational path-integral in mini-superspace approximation)
\beq
\label{eq:Gamp}
G[{\rm bd}_0, {\rm bd}_1]
= \int_{0^+}^{\infty} {\rm d} N_c  \int_{{\rm bd}_0}^{{\rm bd}_1} {\cal D} q(t) \,\, 
\exp \left(\frac{i}{\hbar} S_{\rm tot} \right) \, , 
\eeq
where ${\rm bd}_0$ and  ${\rm bd}_1$ are initial and final 
boundary configurations respectively, and
$S_{\rm tot}$ for the NBC is given in eq. (\ref{eq:Sact_frw_simp_nbc}). 
The path-integral over $q(t)$ is performed while respecting the boundary conditions. 
The original contour of integration for lapse $N_c$ is $(0^+, \infty)$. 

We start by considering the fluctuations around the
solution to equation of motion, which has been obtained previously
respecting the Neumann boundary conditions.  
\beq
\label{eq:qdecomp}
q(t) = \bar{q}(t) +  \ep^\prime Q(t) \, ,
\eeq
where $\bar{q}(t)$ is the solution to equation of motion 
given in eq. (\ref{eq:qsol_nbc}), 
$Q(t)$ is the fluctuation around the background $\bar{q}$, 
and $\ep^\prime$ is a parameter introduced to keep a track of terms.
The decomposition in eq. (\ref{eq:qdecomp}) can be plugged back 
in total action given in (\ref{eq:Sact_frw_simp_nbc}) and expanded in 
powers of $\ep^\prime$. The fluctuation $Q(t)$ obeys a similar 
set of boundary conditions as the background $\bar{q}$:
namely fixing $\dot{Q}_0$ and $Q_1$ at the initial and final 
boundary respectively. This means
\beq
\label{eq:QBC}
{\rm Specify\,\,} \dot{Q}_0 {\rm \,\, and\,\,} Q_1 
\hspace{5mm}
\Rightarrow 
\hspace{5mm}
Q_1 = \dot{Q}_0=0 \, .
\eeq
After imposing these Neumann boundary conditions on $Q$ we 
perform an expansion of action in powers of $\ep^\prime$.
We notice that first order terms in $\ep^\prime$ vanish (as expected) 
as it is proportional to equation of motion for $\bar{q}(t)$. 
The second order terms are non-vanishing. The series in $\ep^\prime$ stops at 
second order and there are no more terms in the series. 
The full expansion can be written as
\beq
\label{eq:Sexp_Q}
S_{\rm tot}= S_{\rm tot}[\bar{q}] 
- \frac{3 \ep^{^\prime2}}{4N_c} \int_0^1 {\rm d}t \dot{Q}^2 \, ,
\eeq
where $S_{\rm tot}[\bar{q}]$ is given in eq. (\ref{eq:stot_onsh_nbc}). 
The path-integral measure after the above decomposition will become the following
\beq
\label{eq:Mes_dec}
\int {\cal D} q(t) \Rightarrow \int {\cal D} Q(t) \, .
\eeq
As the action given in eq. (\ref{eq:Sexp_Q}) separates into two parts: 
a part independent of $Q$ and part quadratic in $Q$, 
therefore one can perform the path-integral over $Q$  
independently of the rest. This path-integral over $Q$ is 
\beq
\label{eq:pathQ_sep}
F(N_c) = 
\int_{Q'[0]=0}^{Q[1]=0} 
{\cal D} Q(t) 
\exp \left(
- \frac{3 i \ep^{^\prime2}}{4 \hbar N_c} \int_0^1 {\rm d}t \dot{Q}^2
\right) \, .
\eeq
This path-integral is very similar to the path-integral for a free fields
with end points field values kept fixed. However, this one is 
a bit different as at the initial boundary we are fixing $\dot{Q}$. 
Following the footsteps in \cite{DiTucci:2020weq} we note
\beq
\label{eq:FNc_mbc}
F(N_c) = \frac{1}{\sqrt{\pi i}} \, .
\eeq
The crucial point to note here is that in the case of 
mixed boundary conditions like the ones considered here,
the above path-integrals gives a 
$N_c$-independent numerical factor. This is unlike the case in 
Dirichlet boundary conditions where the 
path-integral like the one above is proportional to $N_c^{-1/2}$. 
On plugging the expression for $F(N_c)$ we obtain an expression for the 
transition amplitude $G[{\rm bd}_0, {\rm bd}_1]$ where the integration 
limits have been extended all the way to $-\infty$. This means that the 
transition amplitude is given by
\beq
\label{eq:Gab_afterQ}
G[{\rm bd}_0, {\rm bd}_1]
= \frac{1}{2\sqrt{\pi i}} \int_{-\infty}^\infty {\rm d} N_c \,\, 
\exp \left(\frac{i}{\hbar} S_{\rm tot}[\bar{q}] \right) \, ,
\eeq
where $S_{\rm tot}[\bar{q}]$ is given in eq. (\ref{eq:stot_onsh_nbc}). 

To deal with the lapse integration we first make a change of variables.
We shift the lapse $N_c$ by a constant 
\beq
\label{eq:NcTONb}
N_c = \bar{N} + \frac{\pi_0}{\Lam} \, 
\hspace{5mm}
\Rightarrow
\hspace{5mm}
{\rm d}N_c \hspace{3mm} \to \hspace{3mm} {\rm d} \bar{N} \, .
\eeq
This change of variable will imply that the action for the lapse 
$S_{\rm tot}[\bar{q}]$ becomes the following 
\beq
\label{eq:sact_NcTONb}
S_{\rm tot} 
= \frac{\Lam^2}{9} \bar{N}^3 + (3k - \Lam q_1) \bar{N}
+ \left(\frac{3}{\Lam} + 4\al \right) \left(k \pi_0 + \frac{\pi_0^3}{27} \right) \, .
\eeq
It is interesting to note here that after the change of variables the $\pi_0$
dependence only appears in the constant term. After the change of variables 
the transition amplitude is given by
\bea
\label{eq:Gab_Nb}
G[{\rm bd}_0, {\rm bd}_1]
= &&
\frac{1}{2\sqrt{\pi i}} \exp \left[\frac{i}{\hbar} 
\left(\frac{3}{\Lam} + 4\al \right) \left(k \pi_0 + \frac{\pi_0^3}{27} \right) \right]
\notag \\
&& \times
\int_{-\infty}^\infty {\rm d} \bar{N} \,\, 
\exp \left[\frac{i}{\hbar} \left\{\frac{\Lam^2}{9} \bar{N}^3 + (3k - \Lam q_1) \bar{N} \right\} \right] 
= \Psi_1(\pi_0) \Psi_2(q_1)
\, ,
\eea
where 
\bea
\label{eq:psi1}
&&
\Psi_1(\pi_0) = \frac{1}{2\sqrt{\pi i}} \exp \left[\frac{i}{\hbar} 
\left(\frac{3}{\Lam} + 4\al \right) \left(k \pi_0 + \frac{\pi_0^3}{27} \right) \right] \, ,
\\
\label{eq:psi2}
&&
\Psi_2(q_1) = \int_{-\infty}^\infty {\rm d} \bar{N} \,\, 
\exp \left[\frac{i}{\hbar} \left\{\frac{\Lam^2}{9} \bar{N}^3 + (3k - \Lam q_1) \bar{N} \right\} \right] \, .
\eea
The transition amplitude for the case of the Neumann boundary condition 
at the initial boundary and a Dirichlet boundary condition at the 
final boundary is a product of two parts: one 
given by $\Psi_1$ is entirely dependent on initial momentum $\pi_0$ 
and other $\Psi_2$ is function of $q_1$ which tells the final size of Universe. 
The dependence of amplitude 
on two boundary configurations gets separated. 
This kind of factorization was also observed in a recent paper 
\cite{Lehners:2021jmv} where the authors studied the 
Wheeler-DeWitt (WdW) equation in mini-superspace 
approximation of Einstein-Hilbert gravity.
Also note that we are working in the convention $V_3 = 8\pi G$, 
factors of which can be restored later when needed.

\section{Airy function}
\label{airy}

In this section we will study in more detail the nature of $\Psi_2(q_1)$
which is mentioned in eq. (\ref{eq:psi2}). It should be noted that this function 
can be identified with \textit{Airy}-integral. This integral 
also depends crucially on the contour of integration along which 
it has to be integrated. For the case of Airy-function the  
regions of convergence are located at the following phase angles
$\ta \equiv \arg(\bar{N})$: $0\leq \ta \leq \pi/3$ (region $1$),
$2\pi/3 \leq \ta \leq \pi$ (region $0$), and 
$4\pi/3 \leq \ta \leq 5\pi/3$ (region 2). The following 
contours can be defined: ${\cal C}_0$ the contour 
running from region $0$ to region $1$, 
${\cal C}_1$ the contour 
running from region $1$ to region $2$, and 
${\cal C}_2$ the contour 
running from region $2$ to region $0$.
Using these contours one can define the following two types 
of Airy-integrals 
\bea
\label{eq:airy_ai}
&&
Ai(z) = \frac{1}{2\pi} \int_{{\cal C}_0} {\rm d} x \exp
\left[i\left(\frac{x^3}{3} + z x \right) \right] \, , 
\\
\label{eq:airy_bi}
&&
Bi(z) = \frac{i}{2\pi} \int_{{\cal C}_2 - {\cal C}_1} {\rm d} x \exp
\left[i\left(\frac{x^3}{3} + z x \right) \right] \, .
\eea
There are two cases now: $\Lam<0$ (AdS-geometry)
and $\Lam>0$ (dS-geometry). We will see that it is possible 
to analytically continue one to another as the Airy-function 
is an analytic function. We will first look at the case 
when $\Lam<0$ then analytically continue it to $\Lam>0$ 
to obtain the exact result for the dS case.

For $\Lam<0$ we have $3k-\Lam q_1>0$ for all $q_1\geq0$. This will mean 
that the argument of the Airy-functions are real, 
and thus the values of the Airy functions are also real. 
As discussed in \cite{DiTucci:2020weq}, one uses the knowledge 
of expected CFT results for a comparison. This implies two 
things: (1) that the function $\Psi_2$ must be real 
and (2) the function $\Psi_2$ should have volume 
divergence \textit{i.e.} as $q_1\to\infty$ the function should show 
divergence which has to be appropriately removed by addition 
of suitable counter-terms (we will not discuss the computation 
of counter-terms here, see \cite{DiTucci:2020weq}).

\begin{figure}[h]
\centerline{
\vspace{0pt}
\centering
\includegraphics[width=12cm]{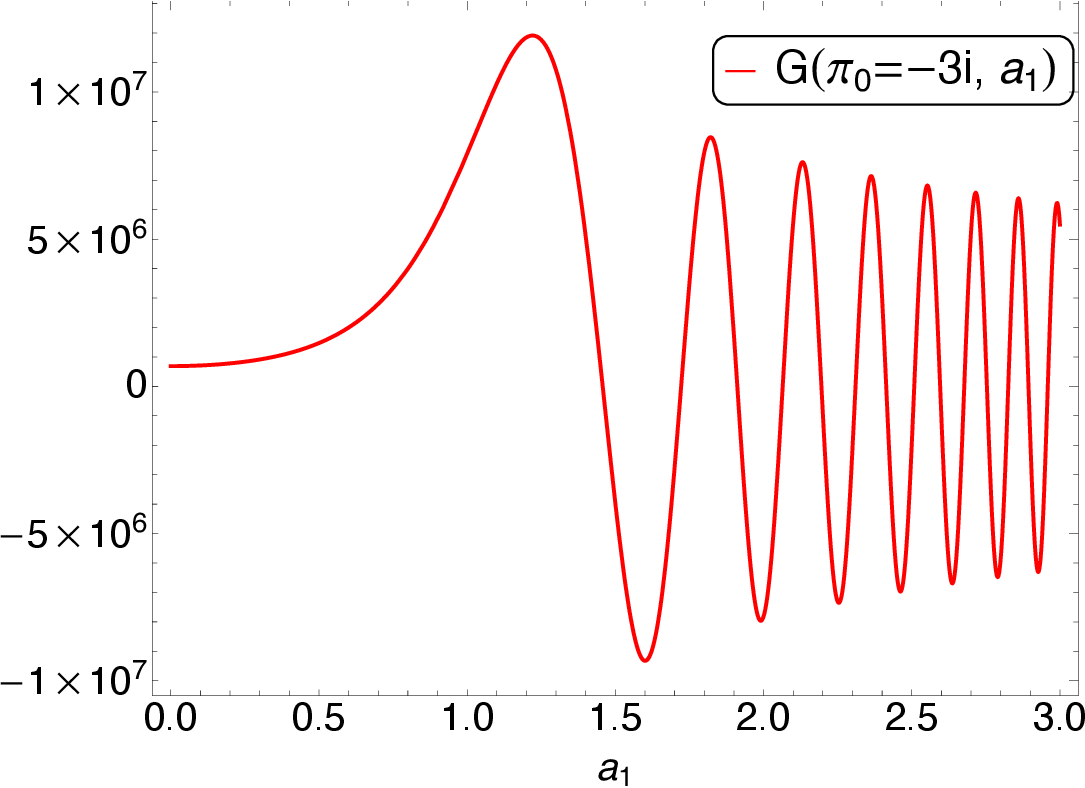}
}
\caption[]{
We consider the boundary conditions motivated by no-boundary Universe which imply 
$\pi_0=-3i$. Here 
we choose parameter values: $k=1$, $\Lam=3$, and $\al=2$. 
Here we plot the 
exact transition amplitude $G(\pi_0=-3i, a_1)$ given in 
eq. (\ref{eq:Gbd0bd1_full_LAMge0}) as $a_1 = \sqrt{q_1}$ is varied
from $0$ to larger values. 
}
\label{fig:gamp_csb}
\end{figure}
%

The asymptotic expressions for the two Airy functions are:
$Ai(z) \sim \exp(-2 z^{3/2}/3)$ and $Bi(z) \sim \exp(2 z^{3/2}/3)$.
This immediately tell us that the relevant Airy-function that we seek 
is $Bi(z)$ as that is the one which grows large when $z$ becomes 
large. This will imply that for $\Lam<0$ we have 
\beq
\label{eq:LamLe0_psi2}
\biggl. \Psi_2(q_1) \biggr|_{\Lam<0}= \left(\frac{24 \pi G \hbar}{V_3 \Lam^2}\right)^{1/3}
Bi \left[
\left(\frac{\sqrt{3} V_3}{-8 \pi G \hbar \Lam} \right)^{2/3} \left(3k - \Lam q_1 \right)
\right] \, ,
\eeq
where we have reinstated factors of $V_3$ and $G$ in the above.
Now to obtain the expression for the case of $\Lam>0$ we can 
do an analytic continuation by making use of the following identity 
relating the Airy-functions $Ai(z)$ and $Bi(z)$
\beq
\label{eq:idenAiBi}
Bi(z) = i \om Ai(\om z) - i \om^2 Ai(\om^2 z) \, ,
\eeq
where $\om = e^{i2\pi/3}$ is the cube-root of unity. 

This means that in our case the two Airy functions satisfy the following relation
\bea
\label{eq:BiAi_rel}
Bi \left[
\left(\frac{\sqrt{3} V_3}{-8 \pi G \hbar \Lam} \right)^{2/3} \left(3k - \Lam q_1 \right)
\right] 
&&
= e^{i\pi/6} Ai\left[ e^{i2\pi/3}
\left(\frac{\sqrt{3} V_3}{-8 \pi G \hbar \Lam} \right)^{2/3} \left(3k - \Lam q_1 \right)
\right] 
\notag \\
&&
+ e^{-\pi/6} Ai\left[e^{-i2\pi/3}
\left(\frac{\sqrt{3} V_3}{-8 \pi G \hbar \Lam} \right)^{2/3} \left(3k - \Lam q_1 \right)
\right] 
\notag \\
&&
= \sqrt{3} Ai\left[
\left(\frac{\sqrt{3} V_3}{8 \pi G \hbar \Lam} \right)^{2/3} \left(3k - \Lam q_1 \right)
\right] \, . 
\eea
We notice that for $\Lam>0$ the function $\Psi_2(q_1)$ is given by
\beq
\label{eq:Lamge0_psi2}
\biggl. \Psi_2(q_1) \biggr|_{\Lam>0}= \sqrt{3}\left(\frac{24 \pi G \hbar}{V_3 \Lam^2}\right)^{1/3}
Ai\left[
\left(\frac{\sqrt{3} V_3}{8 \pi G \hbar \Lam} \right)^{2/3} \left(3k - \Lam q_1 \right)
\right] \, ,
\eeq
which is real. If we combine this with the expression for $\Psi_1(\pi_0)$
then we get the full expression for the transition amplitude 
\bea
\label{eq:Gbd0bd1_full_LAMge0}
\biggl. G[{\rm bd}_0, {\rm bd}_1] \biggr|_{\Lam>0}
= 
&&
\sqrt{\frac{3}{\pi i}} 
\left(\frac{3 \pi G \hbar}{V_3 \Lam^2}\right)^{1/3}
\exp \left[\frac{iV_3}{8\pi G\hbar} 
\left(\frac{3}{\Lam} + 4\al \right) \left(k \pi_0 + \frac{\pi_0^3}{27} \right) \right]
\notag \\
&&
\times 
Ai\left[
\left(\frac{\sqrt{3} V_3}{8 \pi G \hbar \Lam} \right)^{2/3} \left(3k - \Lam q_1 \right)
\right] \, .
\eea
This is an exact result for the transition amplitude for the $\Lam>0$ 
in the case when Neumann boundary conditions are imposed at initial boundary and 
Dirichlet boundary conditions are imposed on final boundary. 
Notice also the non-trivial correction coming from the Gauss-Bonnet 
coupling in the exponential factor. This GB-correction however doesn't 
appear in the argument of the Airy-function, which is 
also independent of initial momentum $\pi_0$.

Notice that so far we haven't assumed any special value for
$\pi_0$, which is chosen by demanding the geometry to 
be non-singular at an initial time and fluctuations to be 
well-behaved. For $k=1$ if $\pi_0$ is such that 
${\rm Im} [\pi_0 (1+ \pi_0^2/27)]>0$ then it will lead to an 
exponential with positive real part, and negative otherwise. 
In the next section we will see there is 
one such $\pi_0$ which fits these criterion and is favourable. 

In figure \ref{fig:gamp_csb} we plot the exact transition amplitude 
for a certain choice of parameter values. It is easy to notice its characteristic 
features: namely it rises exponentially from $q_1=0$ (note that value of 
amplitude at $q_1=0$ is non-zero) to $q_1 = 3k/\Lam$ after which it starts 
to oscillate with increasing frequency but with diminishing amplitude. 
In the next section we will study the saddle-point picture to better 
understand the behaviour of the transition amplitude.

\section{Saddle-point approximation}
\label{sad_pot_approx}

The lapse integration mentioned in eq. (\ref{eq:Gab_afterQ}) can also 
be studied using Picard-Lefschtez technology and evaluating it in the saddle-point 
approximation. Although it doesn't provide us with an exact result but
saddle-point analysis helps us in understanding 
the behaviour of transition amplitude as $q_1$ increases
(see \cite{Feldbrugge:2017kzv, Narain:2021bff, Witten:2010cx, Witten:2010zr,
Basar:2013eka, Tanizaki:2014xba} for review on Picard-Lefschtez and 
analytic continuation). 

We start by first studying the action in eq. (\ref{eq:stot_onsh_nbc})
and computing the saddle points for the lapse $N_c$. 
These can be determined 
by computing the quantity $d S_{\rm tot}/d N_c$. 
These saddle points can be obtained by solving the equation
\beq
\label{eq:sad_pt_eq}
\frac{{\rm d} S_{\rm tot}}{{\rm d} N_c} 
= \frac{\Lam^2}{3} N_c^2 - \frac{2\Lam \pi_0}{3} N_c 
+ 3k - \Lam q_1 + \frac{\pi_0^2}{3} = 0 \, .
\eeq
This is a quadratic equation in $N_c$ resulting in two saddle points. 
The discriminant $\D$ for the above equation is given by
\beq
\label{eq:disc_quad}
\D = 4\Lam^2 \left(\frac{\Lam q_1}{3} - k \right) \, .
\eeq
It is crucial to note here that for $\Lam>0$ (dS geometry) the discriminant $\D$ can change 
sign depending on the value of $q_1$ (size determining the outer boundary).
This will result in stokes phenomena as will be seen later
(in the case of AdS-geometry $\Lam<0$ sign of 
$\D$ never changes). The two saddle points are then given by
\beq
\label{eq:sad_pt_nbc}
N_{\pm} = \frac{3}{\Lam} \left[
\frac{\pi_0}{3} \pm \left(\frac{\Lam q_1}{3} - k \right)^{1/2}
\right] \, . 
\eeq
It is worthwhile to remark here that as expected these saddles don't 
depend GB coupling $\al$ for the case of NBC. The action at these saddle 
points becomes the following
\beq
\label{eq:Sad_act_ncb_nc}
S_{\rm tot}(N_\pm) = \pm \frac{6}{\Lam} \left(\frac{\Lam q_1}{3} - k \right)^{3/2}
+ \left(\frac{3}{\Lam} + 4\al \right)\left(k \pi_0 + \frac{\pi_0^3}{27}\right) \, .
\eeq
Notice that the change of variable mentioned in eq. (\ref{eq:NcTONb}) 
will imply that the saddle points for the shifted lapse $\bar{N}$ are given by,
\beq
\label{eq:sad_pt_nbc_nb}
\bar{N}_{\pm} =  \pm \frac{3}{\Lam} \left(\frac{\Lam q_1}{3} - k \right)^{1/2} \, .
\eeq
Now there are two possibilities: $q_1>3k/\Lam$ and $q_1<3k/\Lam$.
In the former case ($q_1>3k/\Lam$) the saddle points given in 
eq. (\ref{eq:sad_pt_nbc_nb}) lie on real axis ($\bar{N}_+$ on positive real axis 
and $\bar{N}_{-}$ on negative real axis). In the case when $q_1<3k/\Lam$
the saddle points are both imaginary with one lying on positive imaginary 
axis while the other lying on negative imaginary axis. The case when 
$q_1=3k/\Lam$ is degenerate when both saddle points are located 
at $\bar{N}_\pm =0$. These three cases have been studied in detail 
in \cite{Lehners:2021jmv} for the case of Einstein-Hilbert gravity.
In our study we notice that the presence of Gauss-Bonnet coupling 
gives an additional contribution 
but doesn't change the overall qualitative picture as far as saddle-point 
analysis is concerned. 

We will now make a particular choice for the initial momentum 
$\pi_0$ using which will proceed with the saddle-point analysis 
to gain more insight of the nature of transition amplitude and
the evolution of Universe. To do this we take inspiration from 
no-boundary Universe where the intuitive understanding 
is that the geometry gets rounded off at beginning of time. 
This means that if we write $\Lam=3\lam^2>0$ in $d=4$ then 
we have for the Lorentzian spacetime dS metric in eq. (\ref{eq:frwmet})
\beq
\label{eq:atp_ds}
N_p = 1 \, , 
\hspace{5mm}
a(t_p) = \frac{1}{\lam} \cosh \left(\lam t_p\right) \, .
\eeq
deSitter geometry when embedded in $5$-dimensions then in closed slicing it can 
pictured as hyperboloid having a minimum spatial extent at $t_p=0$. 
Now the rounding off the geometry is 
achievable by analytically continuing the original dS-metric 
to Euclidean time, starting exactly at the waist of hyperboloid at $t_p=0$. 
This means 
\beq
\label{eq:eucTp}
t_p = \mp i \left(\tau - \frac{\pi}{2\lam} \right) \, , 
\hspace{5mm}
0 \leq \tau \leq \frac{\pi}{2\lam} \, .
\eeq
This means that along the Euclidean section the dS metric transforms 
in to that of a $4$-sphere 
\beq
\label{eq:4sep}
{\rm d}s^2 = {\rm d} \tau^2 + \frac{1}{\lam^2} \sin^2 \left(\lam \tau \right) {\rm d} \OM_3^2 \, .
\eeq
This geometry has no boundary at $\tau=0$ and smoothly closes off. 

Now there are two possibilities of the time rotation to the above 
Euclidean time. Each corresponding to the sign appearing in eq. (\ref{eq:eucTp})
and each leading to a different Wick rotation. 
The upper sign which is also used in the standard Wick rotation in the flat spacetime QFT, 
is also the sign chosen in the works 
of Hartle and Hawking \cite{Hartle:1983ai,Halliwell:1984eu}. In this sign choice 
perturbations are stable and suppressed. 
The lower sign in eq. (\ref{eq:eucTp}) however 
correspond to Vilenkin's tunneling geometry 
in which perturbation are unsuppressed
\cite{Feldbrugge:2017fcc,Halliwell:1989dy}. The  
Wick rotation process here can also be thought of as 
the lapse $N_p$ changing its value from $N_p=1$ to $N_p = \mp i$, 
thereby implying that the 
total time $T_p = \int N_p {\rm d} t_p$ becoming complex valued. 

This when translated in language of metric given in eq. (\ref{eq:frwmet_changed})
will imply
\beq
\label{eq:HHlapse_tr}
\sinh\left(\lam t_p \right)= \lam^2 N_{\rm HH} t + i \, ,
\eeq
where $N_{\rm HH}$ will turn out to be the saddle-point value of
of the lapse corresponding to Hartle-Hawking geometry \cite{Hartle:1983ai,Halliwell:1984eu}.
It is given by
\beq
\label{eq:HHsad}
N_{\rm HH} = \frac{\sqrt{\lam^2 q_1 - 1}}{\lam^2} - \frac{i}{\lam^2} \, ,
\eeq
If we compare this with the saddle point values mentioned in 
eq. (\ref{eq:sad_pt_nbc}) then for our case this will imply 
that for the no-boundary Universe one has
\beq
\label{eq:HHpi0}
\left. \pi_0 \right|_{\rm Hartle-Hawking \,\,(HH)}= - 3 i \, .
\eeq
For this value of $\pi_0$ the initial geometry is 
smoothly rounded-off and is non-singular. Moreover for 
this value the fluctuations are also well-behaved around the saddle points 
and are suppressed. In the following we will assume this particular 
value for $\pi_0$ to proceed with the saddle-point analysis. 
In this case it is noticed that when this value of $\pi_0$ is
plugged into the lapse action given in eq. (\ref{eq:stot_onsh_nbc})
then the action becomes complex. This complex acton is given by
\beq
\label{eq:Sbact_sp}
S_{\rm tot}^{\rm HH} = \frac{\Lam^2}{9} N_c^3 
+ i \Lam N_c^2 - \Lam q_1 N_c - 3 i q_1 - 8\al i  \, ,
\eeq
while the action at the two saddle-points $N_\pm^{HH}$ is given by 
\beq
\label{eq:Sact_HH_sp}
S^{\rm HH}_\pm =  
\pm \frac{6}{\Lam} \left(\frac{\Lam q_1}{3} - 1 \right)^{3/2}
- 2 i \left(\frac{3}{\Lam} + 4\al \right)
 \, .
\eeq
We can also compute the second-derivative of the lapse-action at the 
saddle-points and this is given by
\beq
\label{eq:2nd_der_lap_act}
\biggl. \frac{{\rm d}^2 S}{{\rm d} N_c} \biggr|_{N_c = N_\pm}
= \pm 2 \Lam \sqrt{\frac{q_1 \Lam}{3} -k} \, .
\eeq
It should be mentioned that the second-derivative is 
independent of initial momentum $\pi_0$, and 
that the saddle-point approximation will work 
as long as ${\rm d}^2 S/{\rm d} N_c$ don't vanish for some value $q_1$.
The complex action in eq. (\ref{eq:Sbact_sp}) is a direct 
consequence of imposing complex initial momentum, which 
subsequently leads to complex geometries. A complex action will imply
that even for geometries with real lapse $N_c$ there will be a 
a non-zero weighting corresponding to them.

\begin{figure}[h]
\centerline{
\vspace{0pt}
\centering
\includegraphics[width=4in]{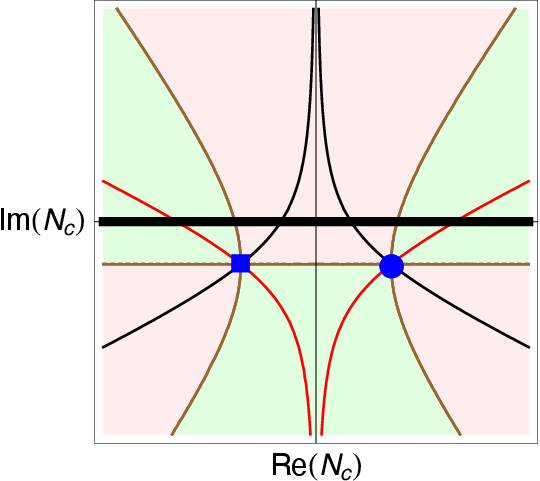}
}
\caption[]{
We consider the case of no-boundary Universe where we impose the mixed 
boundary condition: with Euclidean momentum at $t=0$ and fixed final size at $t=1$. 
The lapse action given in eq. (\ref{eq:stot_onsh_nbc}) is complex. 
We take $\pi_0 = - 3 i$ motivated by works of Hartle \& Hawking 
\cite{Hartle:1983ai,Halliwell:1984eu}. For the purpose of this numerical example 
we have set $\Lam=3$, $\al=2$ and $q_1=3$. 
We plot on $x$-axis real-part of $N_c$ while the $y$-axis is 
imaginary part of $N_c$. 
The red lines correspond to steepest descent lines (thimbles ${\cal J}_\sg$), while 
the thin black lines are steepest ascent lines and denoted by ${\cal K}_\sg$. 
Both the saddle points are depicted in blue: $N_-$ (blue-square) and 
$N_+$ (blue-circle). Both saddle points are {\it relevant}. The 
steepest ascent lines emanating from both of them intersects the 
original integration contour $(-\infty, +\infty)$ which is shown by
thick-black line. The Morse-function $h$ is same for both saddle points:
$h(N_\pm)>0$. The light-green region is allowed region 
with $h<h(N_\sg)$ for all values of $\sg$. The light-pink region 
(forbidden region) has $h>h(N_\sg)$ for all $\sg$. 
The boundary of these region is depicted in brown lines. 
Along these line we have $h = h(N_\sg)$.
}
\label{fig:HH_sp}
\end{figure}
%

At this point our interest turns to compute transition 
amplitude given in eq. (\ref{eq:Gab_afterQ}) by using Picard-Lefschetz 
technology and employing saddle-point approximation. 
Once the saddle points are known, one 
can compute the steepest ascent/descent flow lines corresponding to each of the
saddle point. A {\it relevant} saddle point is one if the steepest ascent path emanating 
from it hits the original integration contour which is $(-\infty, +\infty)$. 
For real action it implies that {\it relevant} saddle points will have 
a negative-valued Morse-function. 
However, when action is complex then this is no longer true
\cite{DiTucci:2019bui,Narain:2021bff,
Lehners:2021jmv, DiTucci:2020weq}.

\subsection{$q_1 > 3/\Lam$}
\label{q1GetLam}

It should be noticed that for $q_1>3/\Lam$ the Morse-function
\beq
\label{eq:Mor_HH}
h(N^{\rm HH}_\pm) = \frac{2}{\hbar} \left(\frac{3}{\Lam} + 4\al \right)\, .
\eeq
It is real, positive and independent of $q_1$. Both saddle points are 
{\it relevant} even though for both of them $h(N_\pm)>0$. 
In figure \ref{fig:HH_sp} we plot the various flow-line, saddle points, forbidden/allowed regions. 
As both saddle points are {\it relevant} so the Lefschetz thimbles 
passing through them constitute the deformed contour of integration. 
The Picard-Lefschetz theory then gives the transition amplitude 
in the saddle point approximation as
\bea
\label{eq:GHH_sp_exp}
&&
G[\pi_0=-3i, q_1>\frac{3}{\Lam}]
\approx \frac{(q_1\Lam/3-1)^{-1/4}}{2\sqrt{2\Lam \pi i}} \left[ 
e^{i\pi/4}\exp\left(\frac{i S^{\rm HH}_{\rm tot}(N_+)}{\hbar}\right)
-e^{-i\pi/4}\exp\left(\frac{i S^{\rm HH}_{\rm tot}(N_-)}{\hbar}\right) 
\right]
\notag \\
&&
= \frac{e^{-i\pi/4}}{(q_1\Lam/3-1)^{1/4}\sqrt{2\Lam\pi}}
\exp\left[
\frac{V_3}{4\pi G\hbar} \left(\frac{3}{\Lam} + 4 \al \right)\right]
\sin \left[\frac{3V_3}{2\pi G \Lam \hbar} \left(\frac{\Lam q_1}{3} - 1\right)^{3/2} + \frac{\pi}{4}\right] \, ,
\eea
where in the last line we have reinstated factors of $V_3$ and $G$.
The transition amplitude oscillates ever faster with increasing $q_1$ 
while its weighting remains constant. This implies that the system 
becomes classical in the WKB sense. Therefore, successive path integrals 
with increasing real boundary values for $q_1$ describe real Lorentzian deSitter 
universes (even though the saddle points in each individual path 
integral have a complex geometry), as long as $q_1 > 3/\Lam$.

\subsection{$q_1 < 3/\Lam$}
\label{q1lessLam}

In this case the discriminant given in eq. (\ref{eq:disc_quad}) is negative,
implying that the term $\left(\Lam q_1/3 - k \right)^{1/2}$ will become 
imaginary. This will mean that the saddle point geometries are 
completely Euclidean in nature. 
The action at the saddle points 
will get an additional imaginary contribution for $\pi_0=-3 i$. 
In this case we notice that the Morse-function $h$ at the saddle points 
changes and is given by
\beq
\label{eq:morse_q1Less1}
h(N^{\rm HH}_\pm) = \pm \frac{6}{\Lam \hbar} \left(1 - \frac{\Lam q_1}{3} \right)^{3/2}
+ \frac{2}{\hbar} \left(\frac{3}{\Lam} + 4\al \right)
\eeq
We therefore realize that only the saddle point $N_{-}$ is relevant. This mean 
that the transition amplitude for $q_1<3/\Lam$ is given by
\bea
\label{eq:GHH_sp_exp_1}
&&
G[\pi_0=-3i, q_1<\frac{3}{\Lam}]
\approx \frac{(1-q_1\Lam/3)^{-1/4}}{2\sqrt{2\Lam \pi i}} 
e^{i3\pi/2}\exp\left(\frac{i S^{\rm HH}_{\rm tot}(N_-)}{\hbar}\right) 
\notag \\
&&
= \frac{e^{-i\pi/2}}{(1-q_1\Lam/3)^{1/4}\sqrt{2\Lam\pi}}
\exp\left[
-\frac{3V_3}{4\pi G\hbar \Lam} \left(1 - \frac{\Lam q_1}{3} \right)^{3/2}
\right]
\exp \left[
\frac{V_3}{4\pi G \hbar} \left(\frac{3}{\Lam} + 4\al \right)
\right]
\, ,
\eea
where in the last line we have reinstated factors of $V_3$ and $G$.
This agrees with the result obtained in \cite{Lehners:2021jmv}
for $\al=0$ case. This also shows that for $q_1<3/\Lam$ the Universe 
is in Euclidean phase.

\subsection{$q_1 = 3/\Lam$}
\label{q1eqLam}

This is a degenerate situation when $\D=0$. In this case 
both the saddle points are same $\bar{N}_\pm=0$. 
The saddle point action is purely imaginary given by
\beq
\label{eq:Sact_HH_q1c}
S^{\rm HH}_\pm (q_1 = \frac{3}{\Lam}) =  
- 2 i \left(\frac{3}{\Lam} + 4\al \right)
 \, ,
\eeq
while the Morse-function is given in eq. (\ref{eq:Mor_HH}).

In this degenerate case the saddle point approximation breaks-down 
as the double-derivative of the lapse action given in eq.(\ref{eq:2nd_der_lap_act})
computed at the saddle-points vanishes. This means that one can't perform
the lapse integration for this degenerate case using saddle-point approximation.
This is a short coming in the saddle point approximation as it cannot be applied in 
such situations. This breakdown of the saddle-point approximation will not depend 
on the value of $\pi_0$. In such a situation one has to look beyond 
saddle-point approximation. The exact result computed in section \ref{airy}
and mentioned in eq. (\ref{eq:Gbd0bd1_full_LAMge0}) doesn't
have this problem and gives a reliable result.

\section{Initial condition independence}
\label{inc_ind}

An interesting feature that is noticed in the above computations is the appearance 
of a hypothetical situation when some of 
the couplings of the gravitational theory 
that is mentioned in eq. (\ref{eq:act}) have a particular relation. 
It is seen that if incase there is a situation when 
\beq
\label{eq:Psi1_van}
\al = - \frac{3}{4\Lam} \, 
\eeq
then the exponential factor appearing in the exact result for the 
transition amplitude given in eq. (\ref{eq:Gbd0bd1_full_LAMge0})
becomes unity as the corresponding 
argument of the exponential function vanishes. 
This happens irrespective of the value of $\pi_0$, although
this doesn't mean that the initial momentum $\pi_0$ is 
arbitrary. 

In the case of saddle-point approximation we notice that when $\al$ and $\Lam$ 
satisfy the relation given in eq. (\ref{eq:Psi1_van}) then 
for $q_1> 3/\Lam$ the Morse-function given in eq. (\ref{eq:Mor_HH})
vanishes. While for $q_1<3/\Lam$ the Morse-function given in 
eq. (\ref{eq:morse_q1Less1}) is left with the part dependent only 
on $q_1$. In both the cases however the second-derivative of 
lapse action given in eq. (\ref{eq:2nd_der_lap_act}) remains 
independent of $\pi_0$ (this is irrespective of the relationship 
between $\al$ and $\Lam$). In both these cases the saddle-point 
approximation will hold as the second-derivative of action at 
the saddle-point don't vanish. However, in both these cases we notice 
that the exponential appearing in the transition amplitude 
in eq. (\ref{eq:GHH_sp_exp} and \ref{eq:GHH_sp_exp_1}) becomes unity,
as once again the corresponding 
argument of the exponential factors vanishes.

\begin{figure}[h]
\centerline{
\vspace{0pt}
\centering
\includegraphics[width=12cm]{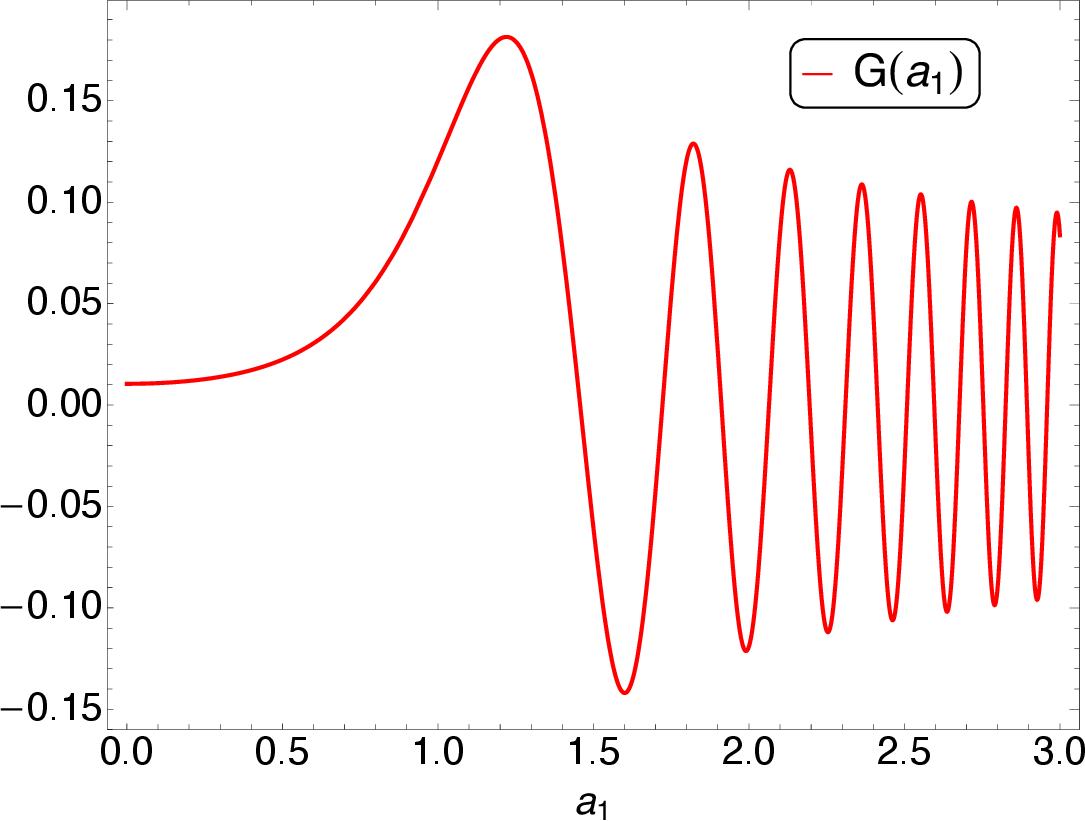}
}
\caption[]{
Here we choose parameter values: $k=1$, $\Lam=3$, and $\al=-\frac{3}{4\Lam}$. 
Here we plot the 
exact transition amplitude $G(a_1)$ given in 
eq. (\ref{eq:sp_wave_nbc}) as $a_1 = \sqrt{q_1}$ is varied
from $0$ to larger values. 
}
\label{fig:gamp_csb_ind}
\end{figure}

It implies that when $\al$ and $\Lam$ satisfy the relation in 
eq. (\ref{eq:Psi1_van}) then we have 
\beq
\label{eq:psi1_1}
\Psi_1(\pi_0) = \frac{1}{\sqrt{\pi i}} \, .
\eeq
At this special value of GB-coupling the dependence on 
the initial momentum $\pi_0$ disappears completely (also note that 
we haven't fix the initial size $q_0$ of the Universe as we were 
considering Neumann boundary conditions at the initial boundary). So for this special 
case the exact transition amplitude (or the wave-function of Universe)
is given by
\bea
\label{eq:sp_wave_nbc}
\biggl. G[{\rm bd}_1] \biggr|_{\Lam>0}
= && \frac{1}{\sqrt{\pi i}} \Psi_2(q_1)
\notag \\
= &&
\sqrt{\frac{3}{\pi i}} 
\left(\frac{3 \pi G \hbar}{V_3 \Lam^2}\right)^{1/3}
Ai\left[
\left(\frac{\sqrt{3 V_3}}{8 \pi G \hbar \Lam} \right)^{2/3} \left(3k - \Lam q_1 \right)
\right] \, .
\eea
This is independent of both the initial size $a_0=\sqrt{q_0}$ (which is arbitrary as we 
imposing Neumann boundary conditions)
and initial momentum $\pi_0$ (which is fixed to some value). 
It only depends on the final size of the Universe 
which is given by $a_1=\sqrt{q_1}$. In figure \ref{fig:gamp_csb_ind}
we plot this transition amplitude as as function of final size of 
Universe given by scale-factor $a_1= \sqrt{q_1}$. 
The qualitative behaviour of transition amplitude as seen from 
figure \ref{fig:gamp_csb_ind} is same as before. For 
$q_1<3/\Lam$ the wavefunction grows exponentially 
indicating a Euclidean phase of Universe when there is 
no notion of time. Note that the value of $G(a_1)$ 
at $a_1=0$ is non-zero although small. 
The wave-function reaches a 
peak value at $q_1=3/\Lam$ and thereafter it has a 
oscillatory feature with diminishing amplitude. This is 
Lorentzian phase of the Universe where emergence of the {\it time} 
has occurred. The system becomes classical in the WKB sense.  

This hypothetical situation of independence from initial 
condition doesn't happen in EH gravity without Gauss-Bonnet modification ($\al=0$ case).
In the case of Gauss-Bonnet gravity the generation of  
extra terms allows for the possibility where terms cancel out.
However, it is important to point out that this independence doesn't 
imply that $\pi_0$ is arbitrary. The initial momentum $\pi_0$ is 
fixed to some value which is chosen by requiring that the 
geometry is smoothly rounded-off at in initial time and 
perturbations are suppressed. Still it is interesting to note 
that the dependence on $\pi_0$ disappears completely 
in the transition amplitude for this hypothetical scenario,
although regularity and stability requirements picks up 
a value for $\pi_0$. 
  
We call this situation `\textit{initial condition independence}'
in a loose sense as the wave-function becomes independent of the 
initial boundary configuration: initial size $q_0$ (which is arbitrary) 
and initial momentum $\pi_0$ (which is fixed in such 
a way so to avoid initial singularity and in which perturbations 
around the saddle-points are suppressed). It should 
be mentioned that for this particular value of $\al$ given in 
eq. (\ref{eq:Psi1_van}) our original action mentioned 
in eq. (\ref{eq:act}) acquires a MacDowell-Mansouri form
\cite{MacDowell:1977jt} in four dimensions 
\beq
S = -\frac{1}{256\pi G} \int {\rm d}^4x \sqrt{-g}
\de^{[\sg \lam \mu \nu]}_{[\g \de \al \bt]}
\biggl(
\sqrt{\frac{3}{\Lam}} R^{\g\de}{}_{\sg\lam}
- \sqrt{\frac{\Lam}{3}} \de^{[\g\de]}_{[\sg\lam]}
\biggr)
\biggl(
\sqrt{\frac{3}{\Lam}} R^{\al\bt}{}_{\mu\nu}
- \sqrt{\frac{\Lam}{3}} \de^{[\al\bt]}_{[\mu\nu]}
\biggr) \, .
\eeq
Interestingly for this same value of $\al$ one also obtains 
finite Noether charges \cite{Miskovic:2006tm}
(see \cite{Miskovic:2009bm} for topological 
regularization). This special value of $\al= - 3/4\Lam$ is
therefore has significance.

\section{Conclusions}
\label{conc}

In this paper we studied the path-integral of the Gauss-Bonnet gravity 
in four spacetime dimensions directly in Lorentzian signature. 
In four spacetime dimensions the Gauss-Bonnet sector of gravity 
is topological in nature and doesn't contribute in the 
bulk dynamics. However, it has an important role to 
play at the boundaries. 
Depending on the nature of boundary conditions the Gauss-Bonnet modifications 
will affect the study of path-integral as has also been noticed in an earlier work 
\cite{Narain:2021bff}. This paper aims to investigate these issues 
in more detail by considering the gravitational path-integral in a reduced setup of 
mini-superspace approximation. 

We start with the mini-superspace action of the theory and vary it with respect 
to the field. This will tell us the dynamical equation of motion and nature of the boundary terms. 
For a consistent variational problem one has to incorporate suitable 
boundary terms. In principle the boundary configurations are chosen 
in such a way so that the variational problem leading to the 
equation of motion (and its solution) is consistent, but it is 
important (and actually better) to choses those BC which leads to 
stable perturbations around the saddle points. 
In these situations the path-integral then reduces to a 
summation over all the \textit{stable} 
geometries, where boundary configurations leading to 
unstable saddles are not incorporated. It is a kind of 
\textit{stability} condition.

We study the system with mixed boundary conditions:
imposing Neumann boundary conditions on the initial boundary and 
imposing Dirichlet boundary condition on the final boundary. 
Motivation for studying mixed boundary conditions stems from various past works:
as the Gauss-Bonnet sector of gravity contributes 
non-trivially \cite{Narain:2021bff}, and imposing 
Neumann BC on the initial boundary is seen to lead 
to saddles where fluctuations are suppressed 
\cite{DiTucci:2019bui,Narain:2021bff,
Lehners:2021jmv, DiTucci:2020weq}. These studies show the 
importance of using Neumann (or Robin) BC at the initial 
boundary. They also further support the situation when
Neumann BC is imposed at initial boundary while a Dirichlet BC is imposed at 
final boundary \cite{DiTucci:2019bui,Narain:2021bff,DiTucci:2020weq, Lehners:2021jmv}, 
as the perturbations are suppressed.

In such a scenario we study the path-integral of 
Gauss-Bonnet gravity in mini-superspace approximation,
and compute the transition amplitude from one 
$3$-geometry to another. This is given by a path-integral 
over $q(t)$ and a contour integration over lapse $N_c$. 
The path-integral over $q(t)$ can be performed exactly 
as the Gauss-Bonnet part only gives some boundary 
contributions. Once this is performed we are left with a
contour integration over the lapse $N_c$
whose action is given by eq. (\ref{eq:stot_onsh_nbc}).

To deal with the lapse integration we do a change of variable 
and shift the lapse by a constant. This allows us to cancel 
some terms in the lapse-action while pushing all the 
dependence on the initial momentum $\pi_0$ in a constant peice. 
This step simplifies the expression for transition amplitude as it 
gets factored into two parts: 
one entirely dependent on the initial momentum 
$\pi_0$ and another which is entirely a function of 
final size of Universe $q_1$. We name these two factors 
$\Psi_1(\pi_0)$ and $\Psi_2(q_1)$ respectively.
The function $\Psi_2(q_1)$ is a contour integral over shifted-lapse
which can be recognized as an Airy-integral. This can be performed 
exactly. We compute this first in AdS-geometry $\Lam<0$ 
(as the argument of the function is positive), then analytically 
continue to the case of dS-geometry ($\Lam>0$) in which we 
are interested. Combining the expression for 
$\Psi_1$ and $\Psi_2$ gives an exact result for the transition 
amplitude in four spacetime dimensions for the case of 
Gauss-Bonnet gravity in mini-superspace approximation. 

We then study the transition amplitude given as a 
lapse integral in eq. (\ref{eq:Gab_afterQ}) in the saddle point 
approximation. We do this to gain more insight in to the behaviour and 
nature of various complex-saddle points as the size of Universe  
$a_1 = \sqrt{q_1}$ increases. We take inspiration from the no-boundary 
proposal of Universe where the spacetime geometry at initial 
time is smoothly rounded-off. This aids us in making an educated 
guess for the initial momentum $\pi_0$, which is the value also 
considered by Hartle-Hawking in their past studies and which is known 
to lead to stable perturbations around {\it relevant} saddle point. 
For this choice of $\pi_0$ in the saddle point approximation we notice that for 
$q_1<3/\Lam$ the saddle-point geometry is Euclidean 
and that the transition amplitude is governed by the 
Euclidean saddle-point. While for $q_1>3/\Lam$
the saddle-point consists of complex conjugate pair and both 
complex-saddles contribute to transition amplitude leading to oscillations.

We come across an interesting hypothetical situation when the cosmological 
constant $\Lam$ and Gauss-Bonnet coupling $\al$ are related
as in eq. (\ref{eq:Psi1_van}). In this case the transition 
amplitude becomes independent of the initial momentum 
$\pi_0$. As the initial size of Universe ($q_0$) was left unspecified, 
so the transition amplitude is completely independent of 
initial size $q_0$ (which is left arbitrary as we are imposing 
Neumann boundary conditions at the initial boundary) 
and initial momentum $\pi_0$. Although the dependence on initial 
momentum disappears from the wave-function, 
but it should be mentioned that the initial momentum is not left
arbitrary. It needs to be fixed to a value which is chosen based 
on regularity and stability requirements. 
It is still interesting 
to note that for this special value $\al = -3/4\Lam$, the 
dependence on conjugate momentum $\pi_0$ disappears from the wave-function. 
We call this hypothetical 
situation \textit{initial-condition independence} in a loose sense. 

This hypothetical situation of the wave-function being independent 
of the initial boundary values don't arise in Einstein-Hilbert 
gravity without Gauss-bonnet (GB) modification and/or when the
GB-coupling don't take such a special value. 
The current works shows the non-trivial contributions that arises 
from the Guass-Bonnet terms in the gravitational action when one 
studies the no-boundary proposal of Universe. Furthermore, 
it highlights the importance of a particular special value of 
$\al$ when the wave-function of the no-boundary Universe 
will enjoy an additional independence from initial boundary values, 
something which was not witnessed before.
Interestingly for this special value of $\al$ the  
original gravitational action acquires a MacDowell-Mansouri form,
and has separately been observed to lead to finite 
Noether charges \cite{Miskovic:2006tm}. 
It would be worth exploring such connections in more 
detail in future.

\bigskip
\centerline{\bf Acknowledgements} 

I am thankful to Jean-Luc Lehners and Nirmalya Kajuri for useful discussions. 
I am thankful to BJUT for kind hospitality and support 
during the course of this work.

%


\end{document}